\documentclass[intlimits,twoside,a4paper]{article}

\usepackage{amsmath,amssymb}
\usepackage{graphicx}

\usepackage[T2A]{fontenc}
\usepackage[cp1251]{inputenc}

\usepackage[eqsecnum]{cmpj2}

\issue{2017}{20}{1}{13804}
\doinumber{10.5488/CMP.20.13804}

\title [Synaptic transmission as a cooperative phenomenon in confined systems]
{Synaptic transmission as a cooperative phenomenon in confined systems}

\author[A.V. Chalyi, A.N. Vasilev, E.V. Zaitseva]{A.V. Chalyi\refaddr{label1}\footnote{E-mail:~avchal@nmu.ua}\,, A.N. Vasilev\refaddr{label2}\footnote{E-mail:~vasilev@univ.kiev.ua}\,, E.V. Zaitseva\refaddr{label1}}
\addresses{
\addr{label1} Department of Medical and Biological Physics, Bogomolets National Medical University,\\ 13 Shevchenko Blvd., 01601 Kyiv, Ukraine
\addr{label2} Department of Theoretical Physics, Taras Shenchenko Kyiv National University,\\ 6 Acad. Glushkov Ave., 03022 Kyiv, Ukraine
}

\authorcopyright{A.V. Chalyi, A.N. Vasilev, E.V. Zaitseva, 2017}
\date{Received January 29, 2017, in final form March 12, 2017}

\begin{document}

\maketitle

\begin{abstract}
In this review paper, the theory of synaptic transmission (ST) was developed and discussed. We used the hypothesis of isomorphism between: (a) the cooperative behavior of mediators --- acetylcholine molecules (ACh) and cholinoreceptors in a synaptic cleft with binding into mediator-receptor (AChR) complexes, (b) the critical phenomena in confined binary liquid mixtures. The systems of two (or three) nonlinear differential equations were proposed to find  the change of concentrations of ACh, AChR complexes, and ferment acetylcholinesterase. The main findings of our study: the linear size of the activation zone was evaluated; the process of postsynaptic membrane activation was described as a cooperative process; different approximations of ACh synchronous release were examined; stationary states and types of singular points were studied for the proposed models of ST; the nonlinear kinetic model with three order parameters demonstrated a strange-attractor behavior.
\pacs 82.40.Ck, 87.10.Ed
\end{abstract}

\begin{flushright}
``Biophysics is a Tower of Babel''

P.G. de Gennes

The Nobel Laureate in Physics (1991)
\end{flushright}

\section{Introduction}

The problems of evolution of living (organic) and non-living (inorganic) nature still attract a great interest and agitate inquisitive human mind. Among these problems, synaptic transmission (ST) and cell-to-cell communication play an absolutely outstanding role. Actually, cell-to-cell communication is of the same fundamental importance in animate nature as intermolecular interaction in non-living nature. It makes it possible to understand the basic principles of a decisive problem, i.e., how our brain works and, in particular, how our nonlinear brain system organizes its thinking function \cite{1,2,3,4}.

The understanding of the processes of cell origin (their typical sizes are about $10^{-6}$~m) came from the study of evolution processes in the living nature at  distances intermediate relative to the farthest boundary of the Universe, i.e., more than 10~billion light years or $10^{25}$~m, and at the smallest distances inside the atomic nucleus, i.e., less than   $10^{-15}$~m. The cells emerged due to the creation of plasmatic membranes, i.e., the process concerned with the so-called self-assembly of amphiphilic molecules of phospholipids in aquatic environment, which was absolutely determined (see appendix~\ref{sec:Apb}).

Generation and propagation of electric potentials is the most important phenomenon in living cells and tissues based on excitation of cells, regulation of intracellular processes, muscular contraction, and nerve system functioning. The origin of electric membrane potentials of two types (in particular, the \textit{resting potential} and the \textit{action potential}) may be explained by a specific feature of distribution and diffusion of ions considering their permeability through a membrane \cite{5,6,7,8,9,10}.

In 1949--1952 Hodgkin and Huxley \cite{1} experimentally and theoretically studied the generation and propagation of the action potential. In 1963 they were awarded, together with Eccles, a Nobel Prize in medicine or physiology ``\dots for their discoveries concerning the ionic mechanisms involved in excitation and inhibition in the peripheric and central portions of the cell membrane''.

The problems of ST and cell-to-cell communication have actually two important aspects: \textit{electrical} and \textit{chemical}. The electrical intercellular interaction aspect is associated with the process of the action potential propagation along the nerve fibers (axons).  The chemical aspect is connected with chemical reactions between neurotransmitters and receptors in synaptic clefts (SC), or synapses separating dendrites of one neuron from other neurons, and in neuro-muscular junctions \cite{1,2,3,4}.

It is known that the body (soma) of a nervous cell (neuron), similarly to any other cells, consists of mitochondria, ribosomes and many other organelles: nucleus with its genetic information; dendrites emitting synaptic information from their own soma and receiving this information from other nervous cells; axons transmitting information to  dendrites of other neurons.

Consider some important characteristics and properties of the cell-to-cell communication and the ST \cite{1,2,3,4,5,6,7,8,9,10,11}:  1) size of a human nervous cell (neuron) --- from 10 to 100~{\textmu}m (the largest size is typical of Betz cells --- giant pyramidal cells discovered in 1874 by a Ukrainian anatomist Volodymyr Betz, who worked at the medical faculty of St. Volodymyr Kyiv University \cite{11}); 2) size of SC (thickness of synapse) --- from 10 to 200~nm; 3) diameter of ion channels in the membrane --- up to 10~nm; 4) axon length --- up to 1~m; width (diameter) of axons --- up to $0.5{-}1$~mm for a giant squid axon; 5) the number of neurons in a human brain --- about $10^{11}$ (a more precise number is 86~billions); 6) the number of contacts (synapses) between dendrites of one neuron with the other neurons --- up to $1.5 \cdot 10^4$; 7) the number of synapses in a human brain --- about $10^{14}{-}10^{15}$; 8) time of a signal exchange  --- about $10^{-3}$~s; 9) the number of channels in the membrane surface --- about $10^2{-}10^3$ per {\textmu}m$^2$; 10) the number of impulses generated by a neuron per 1~second ---  from 1 to~$10^3$; 11) the number of neurotransmitters, say, acetylcholine (ACh) molecules contained in 1~vesicle --- about $10^4$; 12) the number of ACh molecules released simultaneously from the axonal presynaptic terminal with approximately $10^3$~vesicles, under the action of one nerve impulse (action potential) --- about~$10^7$.

In this contribution, we propose a new theoretical approach to the problem of ST, based on the nonlinear kinetic models and the hypothesis of isomorphism between (a) the cooperative behavior of a neurotransmitter-receptor system in the synaptic cleft (SC), (b) the critical phenomena in confined binary mixture near its critical mixing point. This hypothesis was originally proposed in 1989 and was published in \cite{12}. Further, it was developed in the lecture ``Critical phenomena in finite-size systems and synaptic transmission'', presented in 1991 at the ARW NATO ``Dynamic phenomena at interfaces, surfaces and membranes'' and was published in the Les Houches Series in \cite{13}. The preface to this book was written by the Nobel Laureate P.G. de Gennes (1991), with its first phrase being used here as an epigraph.

As an application of the above-mentioned hypothesis to cholinergic synapses, the systems of two (or three) nonlinear differential equations were proposed to examine the change of order parameters such as concentrations of ACh, acetylcholine-receptor (R) complexes (AChR), and acetylcholinesterase (AChE), being a ferment which removes ACh from the R-location area. The main findings of our study were as follows: the linear size of the activation zone was evaluated as the correlation length (CL) for a mediator-receptor mixture (section~\ref{sec2}); the process of the postsynaptic membrane activation was described as a cooperative process (section~\ref{sec3}); the nonlinear kinetic model of ST with three order parameters ACh, AChR and AChE demonstrated a strange-attractor behavior, different approximations for the function of ACh synchronous release were examined, the stationary states and types of singular points were studied for the proposed models of ST (section~\ref{sec4}).

\section{Model of ST as phase transition in ``transmitter-receptor'' system} \label{sec2}

The chemical reactions near the bifurcation (critical) points are of a great importance for the ST mechanisms of nerve impulses.  The phenomena occurring near the points (lines) of phase transitions or boundaries of stability demonstrate a universal behavior in physical, chemical, biological and other systems \cite{14,15,16,17,18,19,20,21,22,23,24,25,26,27,28,34,35,36,37}.  The reason is related to cooperative (synergetic) nature of such events, defined by interaction of fluctuations of the characteristic order parameters, their correlations at large spatial and temporal intervals.

To understand this reason more deeply, let us consider the analogy between the equations of motion for 1) systems described by the fluctuation Hamiltonians with two interacting order  parameters \cite{14,15,16,17,18,19,20,21,22,23,26,27,28}, and 2) nonlinear reaction-diffusion kinetic models \cite{6,13,19,29,30,31,32}. To study such systems undergoing phase transitions and critical phenomena, one usually uses the fluctuation  Hamiltonian \cite{14,15,16,17,18,19}
\begin{equation}
H=\sum_{i}H_{\text{LG}}[\varphi_i]+H_{\text{int}}[\varphi_1,\varphi_2].
\label{2.1}
\end{equation}
Here,
\begin{equation}
H_{\text{LG}}[\varphi_i]=\int\left[\frac{1}{2}a_i\varphi_i^2+\frac{1}{4}b_i\varphi_i^4+\frac{1}{2}c_i(\pmb\nabla\varphi_i)^2\right]\rd V
\label{2.2}
\end{equation}
is the Landau-Ginzburg Hamiltonian, while the second term in (\ref{2.1}) corresponds to the Hamiltonian of interaction between two order parameters $\varphi_i^1$  and  $\varphi_i^2$. Taking into account Hamiltonians (\ref{2.1}) and (\ref{2.2}), one can write kinetic equations, or equations of motion for order parameters
\begin{equation}
\frac {\partial{\varphi_i}}{\partial {t}} = -\Gamma_i \frac{\delta H}{\delta \varphi_i}=-\Gamma_i\left(a_i\varphi+b_i\varphi_i^3-c_i\Delta\varphi_i+\frac{\delta H_{\text{int}}}{\delta \varphi_i}\right),
\label{2.3}
\end{equation}
where $\Gamma_i$  are the Onsager coefficients. Actually, such equations (\ref{2.3}) resemble the kinetic equations of reaction-diffusion kinetic models containing both the diffusion term, being proportional to $\Delta \varphi_i$ with the diffusion coefficient  $D_i=\Gamma_i c_i$, and the non-linearity, connected with the $\varphi_i^3$ and corresponding nonlinear terms in the functional derivative $\delta H_{\text{int}}/\delta \varphi_i$.

It is known that in the vicinity of the critical state, Hamiltonian-like systems demonstrate a universal behavior in the sense that their characteristics do not depend on the particular form of a short-range potential of interparticle interaction. And in the critical state, these systems demonstrate long-range effects that also do not depend on the particular form of interpartical potential. Concerning the ST problem, our principal position, which is based on experimental data, is that the main characteristics of the system do not qualitatively depend on the parameters of the local interaction, and the system demonstrates a cooperative long-range behavior at the regime of its ordinal functioning that could be interpreted as a critical-like state (see additional considerations in the end of section~\ref{sec3}).

Convenient theories of cell-to-cell communication and ST are commonly based on the ideas regarding the chemical intermediates (transmitter agents) securing interaction between two neurons in the synaptic cleft or between the motor neuron and muscle fibre in the neuro-muscular junction.

Here, we shall consider the cholinergic synapse and the corresponding neurotransmitter --- ACh. The sequence of major events in cholinergic synapse are as follows: ACh is synthesized by cholineacetyltransferase and stored in spheroid vesicles in the presynaptic membrane, then ACh molecules are released and, after their diffusion through the synaptic cleft, they react with R --- specific ACh receptors. The binding ACh and R$^*$ into ACh--R$^*$ complexes produces conformal changes in the postsynaptic membrane and, therefore, the change in a membrane electric  potential. Finally, ACh is either inactivated by AChE or is removed by diffusion.

Such a scheme of chemical reactions corresponds to the processes in the cholinergic synapse:
\begin{equation}
\text{ACh} + \text{R} \stackrel{k_1}{\Longleftrightarrow} \text{ACh}-\text{R}^* \stackrel{k_3}{\Longrightarrow} \text{P} + \text{R}^* \stackrel{k_2}{\Longrightarrow} \text{R}\,.
\label{2.4}
\end{equation}
Here, R and R$^*$ are receptors in non-active and active states, respectively; P is a product of ACh destruction by the action of AChE.

The kinetic equations describing, in accordance with (\ref{2.4}), the temporal evolution of the receptor concentration R$^*$ in the active state (variable $x$) and the concentration of ACh--R$^*$ complexes (variable $y$), can be written in the following form \cite{12}:
\begin{equation}
\frac{\rd x}{\rd t} = k_{3}y - k_{2}x,
\label{2.5}
\end{equation}
\vspace{-1mm}
\begin{equation}
\frac{\rd y}{\rd t} = k_{1} [\text{ACh}] ([\text{R}]_{0} - y) - (k_{1} + k_{3}) y,
\label{2.6}
\end{equation}
where [ACh] is concentration of ACh molecules, [R]$_{0}$ is initial concentration of  receptors in non-active state, $k_{i}$  are coefficients of reactions velocities.

It must be stressed that the process of ACh release from vesicles in the presynaptic membrane into the synaptic cleft is a cooperative phenomenon: about $10^{7}$~ACh molecules are simultaneously released under the action of one nerve impulse (spike). Such a synchronous activation of a large zone of receptors by ACh molecules can be treated in detail as the process which is isomorphic to the critical phenomena in finite-size binary liquid mixtures near the critical mixing point \cite{13,14,19,20}.

Here, we begin to study the ST process starting with the simplest model \cite{12} with two order parameters and two kinetic equations~(\ref{2.5}), (\ref{2.6}) and then we continue to correspondingly examine more complicated models with three order parameters (and kinetic equations) in section~\ref{sec4}. It is worthy to mention that the 3rd order parameter associated with the concentration of ferment AChE appears at the last stage of ST process when ACh is inactivated by AChE. Actually, the hypothesis of isomorphism corresponds only to the models of ST process in sections~\ref{sec2}, \ref{sec3}. This problem was investigated in detail in chapter~5 of \cite{29} and was explicitly examined in \cite{30,31,32}.

\textit {\textbf{Correlation function and correlation length in ACh--R$^*$ system in SC.}} The expression for the pair correlation function (CF) $G_{2} (r)$ of the Ornstein-Zernike approximation $G_{2} (r) = A \exp (- r / \xi ) / r$, is valid  for 3-dimensional systems with zero boundary conditions and system linear dimension $L_{0} \gg \xi$, where $\xi$ is the correlation length of order parameter fluctuations. As it is known, the concentration is an order parameter of  binary liquid mixture near the critical mixing point (see, e.g., \cite{19,27}). The CF $G_{2} (r) \rightarrow 0$  if $r = |{r_{1}-r_{2}| \rightarrow \infty} $. The anomalous growth of CL $\xi$  and the long-range behavior of the CF $G_{2} (r) \sim r^{-1}$   takes place at the phase transition or critical points only for spatially infinite systems.

To study the correlation properties of ACh--R$^*$ system, a fundamentally important point is the fact that the SC is a finite-size (confined) system. For obvious reasons, it is clear that the CL  $\xi$ in the synaptic cleft cannot exceed the linear size $L_{0}$ of  a system.  Therefore, an actual problem is to find the correlation properties [CF $G_2 (r)$ and CL $\xi$] of ACh--R$^*$ system inside the spatially confined synaptic cleft.

Such a problem was first solved for a plane-parallel geometry of the synaptic cleft in \cite{16} and then in detail in \cite{28,29,30,31,32,33,34}. The main contribution to the CF is given by the following expression:
\begin{equation}
 G_{2} (\rho, z) = \frac{1}{2 \pi L_0} K_0 \left[\rho \left(\kappa ^2 + \frac{n^2 \pi ^2}{4 L_0 ^2} \right)^{1/2}\right] \cos\left(\frac{\pi z}{2 L_0}\right),
\label{2.7}
\end{equation}
where  $\rho=(x^2 +y^2)^{1/2}$, $K_0(u)$ is the cylindrical Macdonald function, and $\kappa=\xi^{-1}$ is the reverse CL of spatially infinite systems. Correlation function~(\ref{2.7}) demonstrates an oscillatory behavior in the $z$ direction of spatial limitation for plane-parallel layers confirming the theoretical results and computer-simulation studies (see e.g. \cite{35}) for the radial distribution function $g(r)$  in liquids in restricted geometry.

As is seen from~(\ref{2.7}), the pair correlation function $G_2(\rho, z)$  of ACh--R$^*$ system in confined geometry has a non-exponential shape. Therefore, it is natural to determine CL $\xi$ of the order parameter fluctuations in such bounded liquids according to the following relation:
\begin{equation}
\xi=\sqrt{M_2}\,, \qquad M_2=\frac{\int G_2(r)r^2 \rd r}{\int G_2(r) \rd r}\,,
\end{equation}
where $M_2$ is the second normalized spatial moment of the pair correlation function.

In the SC with a plane-parallel geometry with thickness $H=2L_0$, taking into account the formu\-la~(\ref{2.7}) for the CF $G_2(r)$, expression for the Macdonald function $K_{\nu}(u)$  and gamma-function $\Gamma(u)$ \cite{30} \linebreak $\int_{0}^{\infty} x^{\mu} K_{\nu}(ax)\rd x = 2^{\mu - 1} a^{- \mu - 1} \Gamma \big((1+\mu+\nu)/2\big) \Gamma \big((1+\mu-\nu)/2\big)$, one can get such a formula for the CL $\xi$  of concentration fluctuations in the SC:
\[\xi=\left[\left(\kappa ^2 + \frac{\pi ^2}{4 L_0 ^2}\right)^{-1}+\left(1-\frac{8}{\pi ^2}\right)\left( \frac{L_0}{2}\right) ^2\right]^{1/2}.\]
An anisotropic behavior demonstrated by CL $\xi$ is determined by two contributions $\xi = (\xi_{xy}^2 + \xi_{z}^2 )^{1/2}$, where $\xi_{xy} =1/(\kappa^2 + \pi ^2 / 4 L_0 ^2)^{1/2} $ is the CL in the $xy$ plane and  $\xi_z = (L_0/2) (1 - 8/ \pi^2)^{1/2}$ is the CL in the $z$ direction, respectively, parallel (perpendicular) to presynaptic (postsynaptic) membrane. The analysis of the expression for   $\xi$ shows that the maximum value of the CL in the $xy$ plane equals $(\xi_{xy})_{\text{max}}= 4L_0 /\pi$  in the mixing point where $\kappa=0$. Usually, the thickness of chemical synapses is about several dozens of nanometers, so we may assume for definiteness that the membrane thickness $H = 2L_0 \approx  20$~nm. Thus, in the case of chemical synapses, the maximum value of the CL in the $xy$ plane equals: $(\xi_{xy})_{\text{max}} \approx 6.4$~nm, while the CL in $z$~direction equals:  $\xi_{z} \approx 2.2$~nm. In a neuro-muscular junction, the thickness $H$ is higher, reaching 100~nm, i.e., $L_0\approx  50$~nm.  In this case, taking the width of the neuro-muscular junction $2L_0\approx 10^{-7}$~m, one has a maximum value of the CL $\xi_{xy}$: $(\xi_{xy})_{\text{max}} \approx 32$~nm, while the CL $\xi_ z \approx 11$~nm.

It is known that at the end plate of the postsynaptic membrane, about several million receptor mole\-cules are placed with a high packing at a density which is equal approximately to $\rho_{\text{surf.\,density}} \approx 10^4$~molecular receptor per 1~{\textmu}m$^2$.  It makes it possible to evaluate the number $N_{\text{ACh--}\text{R}^*}$ of correlated ACh--R$^*$ complexes in the postsynaptic membrane. For this purpose, let us find the surface area $S_{\text{act}} =\pi d_{\text{act}}^2 /4$  of  the activation zone in the postsynaptic membrane of the chemical synapse with taking into account its maximum linear size --- the diameter $d_{\text{act}}\approx (\xi_{xy})_{\text{max}} \approx  6.4$~nm. As a result, one has $S_{\text{act}}\approx \pi (\xi_{xy}^2)_{\text{max}}/4 \approx 32.2~\text{nm}^2 = 32.2\cdot 10^{-6}$~{\textmu}m$^2$. Then, the number $N_{\text{ACh--}\text{R}^*}$ of correlated receptors or ACh--R$^*$ complexes in the postsynaptic membrane of the chemical synapse equals the product of $\rho_{\text{surf.\,density}}$ and $S_{\text{act}}\,$, i.e., one has $N_{\text{ACh--}\text{R}^*} =  \rho_{\text{surf.\,density}}\times  S_{\text{act}} \approx 10^4$ receptors$/${\textmu}m$^2 \times 32.2 \cdot 10^{-6}~${\textmu}m$^2 \approx 32.2 \cdot 10^{-2}$~receptors. This result definitely shows that there are no correlated receptors in a chemical synapse. Thus, receptors of chemical synapses  work independently, without any correlations with each other.

Another situation is realized in the neuro-muscular junctions. Really, due to a larger maximum value of the CL $(\xi_{xy} )_{\text{max}}\approx 32$~nm,  the surface area $S_{\text{act}}$ of the activation zone in the neuro-muscular junction is 25~times larger and equals $S_{\text{act}}\approx \pi (\xi_{xy}^2)_{\text{max}}/4 \approx  805$~nm$^2 = 8.05\cdot 10^{-4}$~{\textmu}m$^2$. Then, the number $N_{\text{ACh--}\text{R}^*}$ of correlated complexes equals: $N_{\text{ACh--}\text{R}^*}=  \rho_{\text{surf.\,density}}\times  S_{\text{act}} \approx 10^4$~receptors$/${\textmu}m$^ 2 \times 8.05\cdot 10^{-4}$~{\textmu}m$^2 \linebreak \approx 8$, i.e., approximately 8~receptors mutually correlate in the neuro-muscular junction.

\section{Model of postsynaptic membrane activation as cooperative process} \label{sec3}

Here, we consider the model which describes the process of the postsynaptic membrane activation as a cooperative process \cite{37,38,39,40}. Our model accounts for the interaction of ACh with R on the postsynaptic membrane. Due to this interaction, the receptors on the postsynaptic membrane transfer to the activated state which means generating a signal by the postsynaptic membrane neuron. Basing on this model, we calculate the temporal evolution of the amount of the R$^*$ on the postsynaptic membrane. In particular, we analyze how the system responds to the ACh pulse and also how R upon the postsynaptic membrane interacts with the constant amount of ACh. The most notable result of our modelling is that the key factor which determines the activation process is the total amount of ACh injected into the cleft.

\textit {\textbf{Basic equation}}. The key process in passing the signal through a SC is the biochemical reaction that activates the postsynaptic membrane. This provides a pulse for generating a signal at the adjacent neuron. In a simplified form, we can treat the activation process as the binding of ACh with the R on the postsynaptic membrane. Thus, within this approach, we have some R on the postsynaptic membrane that are ready to interact with ACh, and also we have some R$^*$ (it means that they have already interacted with ACh). If we denote $x$  to be the concentration of the R, and $y$ to be the concentration of ACh, then we can write the following equations to determine the dynamics of ACh and R:
 \begin{equation}
\frac{\rd x}{\rd t} = k_1(x_0 - x) - k_2 xy,
\end{equation}
\begin{equation}
\frac{\rd y}{\rd t} = f(t) - k_2 xy.
\end{equation}
Here, the $x_0$  stands for the total concentration of R on the postsynaptic membrane, the coefficients $k_1$ and $k_2$  are phenomenological parameters of the model. The function $f(t)$ determines the intensity of ACh injection into the SC. We also neglect here the diffusion processes within the synapse (but in general case, these effects should also be taken into account). The system must be supplemented with the initial conditions $x(0)=x_0$  and $y(0)=0$ which means that at the beginning, all the receptors are in the non-activated state and there is no ACh in the SC.

After making substitutions  $t=k_1 x_0 t$, $x=x_0 u$ and $y=x_0 v$, we get the following system of differential equations:
 \begin{equation}
\frac{\rd u}{\rd t} = k(1 - u) - uv,
\label{3.3}
\end{equation}
 \begin{equation}
\frac{\rd v}{\rd t} = j(t) - uv,
\label{3.4}
\end{equation}
where $k=k_2/(k_1 x_0)$, and $j(t)=f(tx_0^{-1}k_1^{-1})/(k_1x_0^2)$ is the dimensionless function that determines the intensity of the ACh injection into the SC. Initial conditions in this case are of the form $u(0)=1$ and $v(0)=0$.

The system of the nonlinear equations~(\ref{3.3})--(\ref{3.4}) can be solved numerically if we know the function  $j(t)$ (and the value of the parameter $k$). Next, we will consider some special situations.

\textit {\textbf{Acetylcholine in synaptic cleft.}}  First of all, it is interesting to clear out what happens if ACh appears in the SC. We mean that the function $j(t)$ is zeroth, but the initial concentration $V$  of the ACh is nonzero, that is  $v(0)=V$. In this case, the concentration of R decreases at first and then it goes up to the initial unit value, meanwhile the concentration of ACh goes from the initial value down to zero. The concentration of the R$^*$ changes opposite to the concentration of the R: at first, the concentration of the R$^*$ increases and then it goes down to the zeroth value. This behavior is totally expectable since the presence of ACh causes the interaction with the receptors, so the amount of R$^*$  increase. Due to the deactivation process, the amount of the R$^*$ decreases to the initial zeroth value.

\looseness=-1 More realistic situation is observed when some portion (i.e., an impulse) of ACh arrives at the postsynaptic membrane. This process can be modulated with the nonzero function $j(t)$ in equation~(\ref{3.4}). Unfortunately, we do not know the precise expression for the function  $j(t)$. Nevertheless, theoretical and experimental data testify that this function (in the case when a single ACh impulse arrives) should be strictly localized in time. Thus, as the probe function we take the expression of the following form: $j(t)=A \exp [-a(t-t_0)^2]$, where the $A$ constant determines the ``strength'' of the ACh impulse, the $a$  parameter determines the localization of the impulse and the $t_0$  is the time of the impulse arrival. We will consider this case. In other words, we consider the $j(t)$  function and the initial condition for the amount of the ACh concentration being zero, so we take $v(0)=0$. Increasing the parameter $A$,  leads to an increase of the R$^*$ amount. This is a natural effect since increasing the parameter $A$  (when the other parameters are fixed) means increasing the total amount of ACh which is injected into the SC. The amount of the  R$^*$ also depends on the value of the parameter $a$. As we can see, increasing the parameter $a$  causes a decrease of the amount of the R$^*$ due to a decrease of the total amount of ACh injected into the cleft. Thus, we can expectedly state that varying the parameters $A$  and $a$  causes the change of the amount of the R$^*$. Nevertheless, it is notable that if $A \sim \sqrt{a}$,  then changing the parameter $a$  changes the value of the parameter  $A$, but the total amount of ACh injected into the cleft is still the same. Numerical calculations show that this synchronic change of the parameters $A$  and $a$  does not significantly affect the profile of the R$^*$ curve.

\begin{figure}[!t]
\centerline{\includegraphics[scale=0.4]{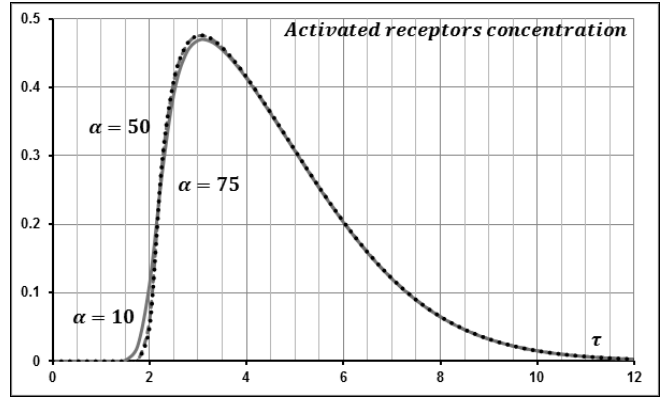}}
\caption{The concentration of the activated receptors as the function of time. It is assumed that  $k=1$, $t_0 = 2$, $A = \sqrt{a}$  and the solid line is for the value  $a=10$, the dashed line is for the value $a=50$  and the dotted line is for the value  $a=75$.}
\label{fig:Fig01}
\end{figure}

Figure~\ref{fig:Fig01} demonstrates the dependence of the concentration of the activated receptors on time. It is obvious that this behavior is universal in the sense that it does not qualitatively depend on the parameter $A = \sqrt{a}$  and, thus, it could be an argument for the initial preposition regarding the isomorphism hypothesis.

This feature can be important for analyzing the ST. Indeed, when studying ACh spreading from the presynaptic membrane through the SC to the postsynaptic membrane, it is important to know the peculiarities of the transportation process and the initial distribution of ACh in the SC. Thus, the question of stability is relative to the ACh distribution profile that could play a crucial role for the ST theory.

\section{Models of ST with M- and N-receptors and effects of diffusion} \label{sec4}
Here, in this section, we shall study the kinetic models of ST with three order parameters and different types of receptors taken into account, specifically, (a) the muscarinergic (M) receptors and (b) the nicotinergic (N) receptors (see appendix~\ref{sec:Apb}).

\textit {\textbf{The 1st kinetic model of ST with 3 order parameters and M-receptors}} takes into account the function of ACh release depending on time, being approximated by using function  $\cos (\Omega t)$, where parameter~$\Omega$  characterizes a slight deviation of the release speed of ACh from the constant value in accordance with the experimental data \cite{42,43,44}. Since the process of release of ACh in the SC is quite fast, in order to take account of this fact and the gradual slowdown supply of ACh molecules in the SC, trigonometric function was expanded in Taylor series preserving the quadratic term in (\ref{4.1}). The corresponding system of differential equations is as follows:
\begin{equation}
\frac{\rd x}{\rd t}=M\left(1-\frac{\Omega^2 t^2}{2}\right)-k_1 x(a-y)-k_2 xz- k_3 x,
\label{4.1}
\end{equation}
\begin{equation}
\frac{\rd y}{\rd t}=k_1 x(a-y)-k_4 xyz,
\label{4.2}
\end{equation}
\begin{equation}
\frac{\rd z}{\rd t}=k_5 z-k_2 xz-k_4 xyz.
\label{4.3}
\end{equation}
The first term in (\ref{4.1}) describes the speed of release of free ACh molecules into SC, the second term describes the rate of binding of ACh with R ($a$ --- total number of M-receptors on the postsynaptic membrane), the third shows the rate of ACh molecules decay at the presence of AChE, the fourth shows the withdrawal of ACh from the synaptic cleft by diffusion and reuptake (return of mediator molecules back to the presynaptic membrane). The following processes are present in (\ref{4.2}): the first term defines a positive contribution to the rate of change of the concentration of ACh--R$^*$ complexes by binding the ACh molecules with R, and the second term describes the speed of the decay of ACh--R$^*$ complexes at the presence of AChE. In (\ref{4.3}), the first term describes the speed of release of free AChE at the decay of ACh--R$^*$ complexes, the second and third terms have the same meaning as similar terms in (\ref{4.1}) and (\ref{4.2}). The stationary system of equations (\ref{4.1})--(\ref{4.3}) has the following solutions:

a) $x_0 = 0$, $y_0 = g$, $z_0 = 0$ ($g <a$). This solution is classified as a strange attractor, because the roots of the characteristic equation are as follows:  $\lambda_1 = - \zeta$, $\lambda_2 = 0$, $\lambda_3 = k_5$, where  $\zeta=k_1 a+k_2 g+k_3$ and $k_5$ is the rate of release of AChE. The emergence of this solution can be explained by the fact that the release function of ACh in the SC is considered as a short, small perturbation;

b) $x_0 = 0$, $y_0 = 0$, $z_0 = h$, where $h$ is the concentration of the enzyme AChE. This solution satisfies system (\ref{4.1})--(\ref{4.3}) provided that $k_5 = 0$. Indeed, for long periods of time, all ACh molecules and ACh--R$^*$ complexes are split and the AChE release rate will be equal to zero. The roots of the corresponding characteristic equation are as follows: $\lambda_1 = - \zeta$, $\lambda_2 = 0$, $\lambda_3 = 0$. In this case, since $\lambda_1 < 0$, $\lambda_2 = 0$, $\lambda_3 > 0$, this singular point is classified as stable torus.

\textit {\textbf{The 2nd model with 3 order parameters and N-receptors}} was created to study the mechanism of ST, which involved N-cholinergic receptors (see appendix~\ref{sec:Apb}). Let us present the processes that occur in the synaptic cleft, with the following scheme of biochemical reactions:
\begin{equation}
\text{ACh} + \text{R} \stackrel{k_1}{\Longleftrightarrow} \text{AChR},
\label{4.4}
\end{equation}
\begin{equation}
\text{AChR} + \text{ACh} \stackrel{k_4}{\Longleftrightarrow} \text{ACh}_2 \text{R} \stackrel{k_4}{\Longleftrightarrow} \text{ACh}_2 \text{R}^*  \stackrel{k_5}{\Longrightarrow} \text{R}.
\label{4.5}
\end{equation}
In other words, at secretion of ACh molecule into the SC being bound to the R (\ref{4.4}), this complex binds with another ACh molecule, then the mediator-receptor complex enters an excited state, and then it returns to the initial state (\ref{4.5}). As in previous models, the reverse processes are not taken into account. Since the activation of N-cholinergic receptors requires two molecules of the ACh, there is an intermediate inactive ACh--R complex with one ACh molecule. To adequately describe such a system, one should increase the number of variables which leads to complications in kinetic equations and increases their numbers. To avoid this, we assume the concentration of the enzyme AChE to be constant. Thus, we will have the following three nonlinear differential equations for the concentration of ACh molecules ($A$), receptors in the unexcited state ($X_1$) and excited mediator-receptor complexes ($X_2$):
\begin{equation}
\dot {A}= M\delta (t)- k_1 AX_1 - k_2 AE - k_3 A - k_4 A[X_0 - (X_1 +X_2)] + k_5 X_2+ k_6 [X_0 - (X_1 +X_2)],
\label{4.6}
\end{equation}
\begin{equation}
 \dot {X_1}= k_5 X_2 - k_1 AX_1 + k_6 [X_0 - (X_1 +X_2)],
 \label{4.7}
\end{equation}
\begin{equation}
\dot {X_2}= k_4 A[X_0 - (X_1 +X_2)] - k_5 X_2\,.
\label{4.8}
\end{equation}
In the first equation, it is accepted that the concentration $E$ of the enzyme AChE is constant, and the release of ACh in the SC is described by Dirac $\delta$-function [see the first term of (\ref{4.6})]. The second and fifth terms describe the binding of the ACh with free R and inactive neurotransmitter-receptor complex, respectively, the third term describes the hydrolysis of ACh in the presence of AChE, the fourth term shows the removal of ACh from the cleft by diffusion and reuptake, and the sixth and seventh terms show disconnection of ACh molecules from the active and inactive neurotransmitter-receptor complex, respectively. Terms, which are included in~(\ref{4.7}): the first and the third mean the same as the sixth and the seventh terms in (\ref{4.6}), respectively, and a free ACh binding with R is described by the second term. In~(\ref{4.8}), the first term describes the transition of an inactive neurotransmitter-receptor complex in the excited state, and the second shows the collapse of neurotransmitter-receptor complex.

The stationary system of equations has a solution: $a=A/A_0=0$, $x_1=X_1/X_0=1$, $x_2=X_2/X_0=0$ ($A_0$ is the total concentration of ACh molecules at the time of release, $X_0$ sets the total concentration of cholinergic receptors in the SC), which corresponds to the real situation when there are no molecules of ACh in the SC, and no mediator-receptor complexes, and all receptors are free. This solution is classified as a stable node because numerical calculation of the roots of the characteristic equation with using experimental data \cite{42,43,44} gives the following result:  $\lambda_1 \approx -5$~ms$^{-1}$, $\lambda_2 \approx -3082$~ms$^{-1}$,  $\lambda_3 \approx -37$~ms$^{-1}$. The corresponding relaxation times are equal to: $\tau_1 \approx 0.2$~ms,  $\tau_2 \approx  3 \cdot 10^{-4}$~ms,  $\tau_3 \approx  2.7 \cdot 10^{-2}$~ms. Since these relaxation times differ quite significantly ($\tau_1 / \tau_2 \approx 670$,    $\tau_1 / \tau_3 \approx 7.5$), the main contribution to the kinetics of the system approach to equilibrium gives the slowest relaxation process with time 0.2~ms.

\textit {\textbf{Kinetic models of ST with 3 order parameters and diffusion effects.}} Here, we have also considered non-linear kinetic models of synaptic transmission in the synapses with M-cholinergic receptors and N-cholinergic receptors, where the impact of diffusion processes was considered more consistently. For both models, it is believed that the process of synaptic transmission has the following common features:

1) molecules of ACh are released in the space of SC very quickly, so the release of ACh function is approximated using the Dirac $ \delta$-function;

2) ACh molecules then diffuse to the postsynaptic membrane, where they bind with the receptor molecules, which causes excitation of nerve impulses in the next cell;

3) then, the mediator-receptor complexes decompose.

In the \textit {\textbf{3rd kinetic model}}, the processes occurring in the cleft of the synapse with M-cholinergic receptors are taken into account. It is assumed that one ACh molecule activates the receptor, and AChE participates in the destruction of neurotransmitter-receptor complexes. The corresponding system of nonlinear differential equations for concentrations of molecules of ACh ($x$), for the ACh--R$^*$ complexes ($y$) and for the enzyme AChE ($z$) is as follows:
\begin{equation}
\dot {x}=K \delta (t)-k_1 x(a-y)-k_2 xz- k_3 x + D_1 \Delta x,
\label{4.9}
\end{equation}
\begin{equation}
\dot {y}=k_1 x(a-y)-k_4 xyz,
\label{4.10}
\end{equation}
\begin{equation}
\dot {z}=k_5 z-k_2 xz-k_4 xyz - D_3 \Delta z.
\label{4.11}
\end{equation}
The first term on the right-hand side of the equation~(\ref{4.9}) describes the rate of ACh release in the SC, the latter shows the removal of neurotransmitter molecules from the cleft by diffusion. The first term in the third equation depending on the sign will have a double meaning: for $k_5 \geqslant 0$ --- a rate of release of the molecules of the enzyme AChE; for $k_5 < 0$,  it is a speed of binding of the enzyme AChE molecules with OPhC that suppresses the speed of release of enzyme molecules. The last term of equation~(\ref{4.11}) describes the movement of molecules of the enzyme in the region with high concentration of acetylcholine through lateral diffusion. All other terms in the equations~(\ref{4.9})--(\ref{4.11}) have the same meaning as the corresponding equations in~(\ref{4.1})--(\ref{4.3}).

Singular points of the system~(\ref{4.9})--(\ref{4.11}):

(a) $x_0 = 0$, $y_0 = g$, $z_0 = 0$, where  $g$ is some concentration of R$^*$, it is clear that $g \leqslant a$;

(b)  $x_0 = m$, $y_0 = a$, $z_0 = 0$, $m$ is some free concentration of ACh;

(c)  $x_0 = 0$, $y_0 = g$, $z_0 = h$,  $h$ is certain concentration of AChE.

Depending on the sign of the parameter $k_5$  that plays the role of a control parameter of the model, points (a), (b) and (c) have the Lyapunov classification (to assess their respective solutions of the characteristic equation, experimental data from the articles \cite{42,43,44} were used):
\begin{itemize}
\item for $k_5 \geqslant 0$  point (a) is classified as a strange attractor, point (b) may be a stable node and a saddle, and point (c) is classified as a strange attractor;
\item for $k_5 < 0$  point (a) is classified as a limiting cycle, point (b) is classified as a stable node, and point (c) is classified as a limiting cycle.
\end{itemize}
It should be noted that lateral diffusion of molecules of the enzyme has less impact on the process of synaptic transmission than the diffusion of mediator molecules.

In the last \textit {\textbf{4th model}},  the influence of the diffusion of mediator molecules on the synaptic transmission was analyzed in the cleft with N-cholinergic receptors. Therefore, a proper system of nonlinear differential equations of the model is substantially similar to the system of equations (\ref{4.6})--(\ref{4.8}) of the second model, but differs from it by the presence of diffusion term $\Delta A$  in the first equation.

Stationary system of equations of the model has the same solution as the 2nd model:  $a=0$, $x_1 = 1$, \linebreak $x_2 = 0$, which corresponds to the real situation when there are no molecules of acetylcholine in the synaptic cleft, and no mediator-receptor complexes, and the receptors are all free.

An analysis of the solutions of characteristic equation shows that at any values of the diffusion coefficient $D$, a singular point is classified as a stable node (the numerical evaluation of experimental data was taken from \cite{42,43,44}) because the roots of the characteristic equation are equal to:  $\lambda_1 \approx -14$~ms$^{-1}$, $\lambda_2 \approx  -3717$~ms$^{-1}$,   $\lambda_3 \approx -37$~ms$^{-1}$. The diffusion process does not change the type of a singular point, although it affects the relaxation time values: $\tau_1= 0.07$~ms, $\tau_2 = 2.6 \cdot 10^{-4}$~ms, $\tau_3 = 2.7 \cdot 10^{-2}$~ms. Having compared the estimates given in the relevant 2nd model, we see that at a consistent account of the processes of diffusion, the relaxation times $\tau_1$ and $\tau_2$ decreased approximately $2.8$ and $1.2$ times, while $\tau_3$ did not change.

\section{Conclusions}

We have created and studied the non-linear kinetic models of ST and intercellular interaction based on actual processes that occur in the synapse of chemical type and isomorphism of coherent aspects of ST and critical phenomena in liquid systems with restricted geometry. The main results and conclusions can be summarized as follows.

1. The linear size of the activation zone was evaluated as the CL $\xi$  for a binary mediator-receptor mixture. It was found that approximately 8~receptors  mutually correlate in the neuro-muscular junction with its thickness of 100~nm.

2. The process of the postsynaptic membrane activation was described as a cooperative process. The system of nonlinear equations for concentrations of R and ACh was numerically solved to find its temporal evolution in different specific conditions. The question of stability relative to the ACh distribution profile could play a crucial role for the ST theory.

3. The analysis of ST models with three order parameters and diffusion effects taken into account showed that if the control parameter changed its sign from positive to negative, the type of singular points also changed --- they all got stability, i.e., the bifurcation occurred: a strange attractor turned into the limiting cycle.

4. Based on the experimental data and numerical calculations for kinetic models with N-receptors, the relaxation times of three order parameters were equal to: $\tau_1= 0.2$~ms,   $\tau_2 = 3 \cdot 10^{-4}$~ms,   $\tau_3 = 2.7 \linebreak\times 10^{-2}$~ms. Thus, the approach of the system to an equilibrium state was characterized by the slowest relaxation process with time $0.2$~ms. Diffusion processes did not change the types of singular points, but affected the relaxation times   $\tau_1= 0.07$~ms, $\tau_2 = 2.6 \cdot 10^{-4}$~ms, while $\tau_3$ did not change.

\section*{Acknowledgements}
The authors wish to thank Professor Yu.V.~Holovatch, a famous scientist in the field of nanobiophysics, critical phenomena and physics of complex systems, one of the best representatives of the widely-known scientific school by Academician I.R.~Yukhnovskii.

\appendix
\section{Abbreviations}\label{sec:Apa}

ST --- synaptic transmission; ACh --- acytineholine; R --- non-activated receptors; R$^*$ --- activated receptors; SC --- synaptic cleft; AChR --- mediator-receptor complexes; AChE --- acetylcholinesterase;  CF --- correlation function;  CL --- correlation length; N --- nicotinergic (receptors); M --- muscarinergic (receptors); OPhC --- organophosphate compounds

\section{Glossary}\label{sec:Apb}

 \textit {Amphiphilic} molecule possesses both hydrophilic and hydrophobic properties. This sort of molecules can form bilayers and micelles in water, so that their polar hydrophilic heads turn to the water molecules, while their hydrophobic tails hide inside the formation.

 \textit {Phospholipids} are a sort of amphiphilic molecules.

 \textit {Mediator} is a biologically active chemical substance transmitting the nerve impulse from one cell to another.

 \textit {Receptor} provides transformation of impacts of environment and inner mediuminto nerve impulse. There are a lot of types of receptors in the human body.

 \textit {Acetylcholine} is a neuromediator providing nerve-to-muscle transmission of the nerve impulse. Without going into biophysical details of intercellular interactions \cite{1,2,3,4,5,6,7,8,9,10}, let us briefly note that there are two types of ACh receptors: 1) only one mediator molecule is sufficient for activation receptors of the first type (so-called \textit {muscarinergic receptors}, or \textit {M-receptors}), while 2) activation of receptors of the second type (so-called \textit {nicotinergic receptors}, or \textit {N-receptors}) requires two mediator molecules. Here, we shall consider several nonlinear kinetic models with M- and N-receptors and the effects of diffusion \cite{40,41}.

 \textit {Acetylcholinesterase} is a hydrolytic ferment catalizing the hydrolysis of acetylcholine which lets the cell change to the rest state.

 \textit {Organophosphor compounds} inhibit the action of \textit {acetylcholinesterase} which leads to the effects of strong poisoning and even to the death of a human body.

\ukrainianpart

\title{Синаптична передача як кооперативне явище в просторово обмежених системах}

\author{О.В. Чалий\refaddr{label1}, О.М. Васильєв\refaddr{label2}, О.В. Зайцева\refaddr{label1}}
\addresses{
\addr{label1} Кафедра медичної та біологічної фізики, Національний медичний університет імені О.О. Богомольця, бульв. Шевченка, 13, 01601 Київ, Україна
\addr{label2} Кафедра теоретичної фізики, Київський національний університет імені Тараса Шевченка,\\ пр. Глушкова, 6, 03022 Київ, Україна
}

\makeukrtitle

\begin{abstract}
\tolerance=3000%
У цьому огляді була розроблена і обговорена теорія синаптичної передачі (СП). Ми застосували гіпотезу ізоморфізму між: (а) кооперативною поведінкою медіаторів --- молекул ацетилхоліну (ACh) і холінорецепторів в синаптичній щілині з утворенням медіатор-рецепторних (AChR) комплексів, (б) критичними явищами в просторово обмежених бінарних рідких сумішах. Були запропоновані системи двох (або трьох) нелінійних диференціальних рівнянь для опису зміни концентрацій ACh, комплексів AChR і ферменту ацетилхолінестерази. Основні результати нашого дослідження: оцінено лінійний розмір зони активації; описано процес активації постсинаптичної мембрани  як кооперативний процес; розглянуто різні апроксимаційні функції синхронного вивільнення ACh; досліджено стаціонарні стани і типи особливих точок для запропонованих моделей СП; показано, що нелінійна кінетична модель з трьома параметрами порядку демонструє поведінку дивного атрактора.
\keywords міжклітинна взаємодія, синаптична передача, фазовий перехід, рідка бінарна суміш, критична точка змішування-розшарування, просторово обмежені системи

\end{abstract}

\end{document}